\documentclass[twocolumn,preprintnumbers,amsmath,amssymb]{revtex4}
\usepackage{graphicx}
\usepackage{dcolumn}
\usepackage{bm}
\usepackage[usenames]{color}
\usepackage{color}
\def\>{\rangle}\def\<{\langle}
\def\togli#1{}

\def\labell#1{\label{#1}}
\def\togli#1{}

\begin{document}
\title{Quantum Time: experimental multi-time correlations}
\author{Ekaterina Moreva$^{1}$, Marco
  Gramegna$^1$*, Giorgio Brida$^1$, Lorenzo Maccone$^3$, Marco Genovese$^{1,4}$}
\affiliation{\vbox{$^1$INRIM, strada delle Cacce 91, 10135 Torino,
    Italy} \vbox{$^3$Dip.~Fisica ``A.~Volta'', INFN Sez.~Pavia, Univ.~of Pavia, via Bassi 6, I-27100 Pavia, Italy}\vbox{$^4$ INFN Sezione di Torino, via P.Giuria 1, 10125 Torino} \vbox{$^{*}$corresponding author: m.gramegna@inrim.it}}

\begin{abstract}
  In this paper we provide an experimental illustration of Page and Wootters' quantum
  time mechanism that is able to describe two-time quantum correlation
  functions.  This allows us to test a Leggett-Garg inequality,
  showing a violation from the ``internal'' observer point of view.
  The ``external'' observer sees a time-independent global state.
  Indeed, the scheme is implemented using a narrow-band single photon
  where the clock degree of freedom is encoded in the photon's
  position. Hence, the internal observer that measures the position
  can track the flow of time, while the external observer sees a
  delocalized photon that has no time evolution in the experiment
  time-scale.
\end{abstract}

\maketitle
\section{Theory}The description of time in quantum mechanics and in particular in connection with quantum gravity and cosmology has always presented significant difficulties \cite{qgt}.
 Page and Wootters (PW) proposed an ingenious framework
\cite{paw} to allow the introduction of a quantum operator for time
that does not exhibit the problems of conventional quantizations of
time, such as the Pauli objection \cite{pauli}. Similar ideas have
appeared also many other times in the literature \cite{morse,altri},
notably in a proposal by Aharanov and Kaufherr \cite{ak}. These
proposals received criticisms \cite{kuchar,unruhwald,sorkin} that were
recently overcome \cite{qtime}. In this last paper, a complete review
of the PW mechanism is presented. For the current aims,
we can summarize their proposal as follows.

In order to quantize time, one can simply define time as ``what is
shown on a clock'' and then use a quantum system as a clock. If time
is to be a continuous degree of freedom, then the clock must be a
continuous system (the position of a photon along a line in our
experimental realization). The requirement that the quantum states of
the system (excluding the clock) satisfy a Schr\"odinger equation
places a strong constraint on the global state of system plus clock:
it must take the entangled form \cite{paw}
\begin{align}
|\Psi\>\>=\int_{-\infty}^{+\infty}dt\:|{t}\>_c|\psi({t})\>_s =
\int_{-\infty}^{+\infty}dt\:|{t}\>_cU_t|\psi({0})\>_s
\labell{paw}\;,
\end{align}
where we use the double-ket notation $|\;\>\>$ to emphasize that the
global state is a bipartite state of clock $c$ and system $s$, and
where $|{t}\>_c$ is the position eigenstate relative to the clock
showing time $t$, $U_t$ is the system's unitary time evolution
operator, and $|\psi({t})\>_s$ is the state of the system at time $t$.
In this framework it is a conditioned state: the state of the system
{\em given} that the clock shows $t$ (a conditional probability
amplitude). The reason for the form of the state $|\Psi\>\>$ in
\eqref{paw} is that one requires that the global state of system plus
clock is time-independent: the system evolves with respect to the
clock and viceversa, so that a global time evolution of system plus
clock would be unobservable. Hence, the global state $|\Psi\>\>$ is a
total energy eigenstate, as in the Wheeler-de Witt equation
$H_g|\Psi\>\>=0$, where $H_g=H_c+H_s$ with $H_g$, $H_c$ and $H_s$ the
global, clock and system Hamiltonians. The Schr\"odinger equation then
follows by choosing a clock that evolves in a ``uniform'' manner
without wavepacket spread, namely with a Hamiltonian $H_c$
proportional to the clock's momentum $\Omega$. Indeed, the
Schr\"odinger equation follows immediately by writing the Wheeler-de
Witt equation in the clock's ``position'' representation:
\begin{align}
{}_c\<{t}|\hbar\Omega+H_s|\Psi\>\>=0\Rightarrow(-i\hbar\tfrac{\partial}{\partial
  {t}}+H_s){}_c\<{t}|\Psi\>\>=0
\labell{wdw}\;,
\end{align}
where $|\psi({t})\>_s={}_c\<{t}|\Psi\>\>$ and where
$-i{\partial}/{\partial {t}}$ is the position
representation of the momentum $\Omega$.

While in \cite{esperimentotorino} we had already presented an
experimental realization of the PW mechanism, the clock
system there (the polarization of one photon) was too simple to allow
for illustrating any but the very simplest features of the mechanism.
Moreover, a two-dimensional clock implies that the time is discrete,
periodic and can take only two values: 0 and 1.  It is then possible
to perform only measurements at two times, namely, to recover only a
single two-time correlation.

In this paper we use a continuous system (the position of a photon) to
describe time, which gives us access to measurements at arbitrary
times and hence arbitrary two-time correlations. We show that one
obtains the correct quantum correlations and we use this to test the
Leggett-Garg inequality \cite{lg} (which, in this framework, was
suggested in \cite{indiani}). One of the main criticisms of the PW mechanism was that it seemed unable to provide the
correct two-time correlations \cite{unruhwald,kuchar,pagereply}. The
main result of the current paper is to experimentally prove that the
mechanism can indeed provide them.

To obtain two-time correlations, we need to be able to describe a
measurement performed at some (internal) time. We will use von
Neumann's prescription for measurements \cite{vonneumannbook}: a
measurement apparatus essentially consists in an (ideally
instantaneous) interaction $U_{vn}$ between the system and a memory
degree of freedom $m$, i.e.~the pointer which stores the outcome.
This interaction is engineered  to correlate the system and the
memory along the eigenbasis $\{|a\>\}$ of the observable $A$ to be
measured, namely
\begin{align}
|\psi({t})\>_s|r\>_m\stackrel{U_{vn}}\longrightarrow
\textstyle{\sum_a}\psi_a({t})|a\>_s|a\>_m
\labell{vn}\;
\end{align}
where $|r\>_m$ is the initial state of the memory, and
$\psi_a({t})=\<a|\psi({t})\>$ is the probability amplitude of
obtaining the $a$-th outcome when measuring $A$. This prescription
gives the expected outcome probabilities $|\psi_a({t})|^2$ through the
Born rule postulate. Writing this measurement evolution in the form of
Eq.~\eqref{paw}, we find \cite{qtime}
\begin{eqnarray}
  &&|\Psi\>\>=\int_{-\infty}^{{t}_a}dt\:
|{t}\>_cU_t|\psi(0)\>_s|r\>_{m}+
\labell{meas}\;\\&&\nonumber
\int_{{t}_a}^{+\infty}\!dt\:|{t}\>_c\sum_a
U_{t-{t}_a}|a\>_s\<a|\psi({t}_a)\>_s
|a\>_{m},
\end{eqnarray}
where ${t}_a$ is the time at which the measurement is performed. The
first integral describes the system evolution prior to the measurement
when the memory is in the $|r\>_m$ state, the second integral
describes the evolution after the measurement, when the memory is now
correlated to the system. Eq.~\eqref{meas} gives the prescription for
a von Neumann measurement in the PW framework (it can
also be extended to general positive operator-valued measures, see
\cite{qtime}).

Superficially, it might seem that the state $|\Psi\>\>$ of
Eq.~\eqref{meas} has lost its time independence, because it depends
explicitly on the measurement time ${t}_a$. However, this parameter
has a meaning as a time only when referred to the internal clock.
Indeed it is a parameter of the global unitary evolution $U_G$ that
contains also the von Neumann interaction: it represents the time
(according to the internal clock) at which the instantaneous
interaction is switched on. It is just a parameter of the total
evolution. More rigorously: using conventional tricks \cite{morse} one
can convert the time-dependent evolution into a time independent one
described by a unitary operator $U_G$ and the evolution \eqref{meas}
becomes $|\Psi\>\>=\int dt\:|t\>_cU_G(t)|\psi(0)\>_s$, where the
dependence on ${t}_a$ has been absorbed into $U_G$. The fact that
\eqref{meas} can be written as a sum of distinct integrals with
different integrands is a consequence of the limit of instantaneous
measurement interaction intrinsic in von Neumann's prescription. In
essence, the fact that the time dependence ${t}_a$ in \eqref{meas} is
referred to the internal clock is clear if one shifts the internal
time axis by an interval $T$. As expected, such translation is
irrelevant to the form of the state \eqref{meas}: it amounts to an
irrelevant change in the integration variable.

If one performs two measurements, then the obvious extension of
\eqref{meas} is
\begin{widetext}
\begin{eqnarray}
  &&|\Psi\>\>=\int_{-\infty}^{{t}_a}dt\:
|{t}\>_cU_t|\psi(0)\>_s|r\>_{m_1}|r\>_{m_2}+
\labell{psi}\;\\&&\nonumber
\int_{{t}_a}^{{t}_b}\!\!\!dt\:|{t}\>_c\sum_a
\<a|\psi({t}_a)\>_sU_{{t}-{t}_a}|a\>_s
|a\>_{m_1}|r\>_{m_2}\!\!+\!\!\!
\int_{{t}_b}^{+\infty}\!\!\!dt\:|{t}\>_c\sum_{ab}
\<b|U({t}_b-{t}_a)|a\>_s
\<a|\psi({t}_a)\>_s
U_{{t}-{t}_b}|b\>_s|a\>_{m_1}|b\>_{m_2},
\end{eqnarray}
\end{widetext}
where $s,c,m_1,m_2$ represent the degrees of freedom of the system,
the clock and the memories of the first and second measurement
apparatus, $U$ is the free unitary evolution operator of the system
(excluding the measurement interaction), $|a\>$ and $|b\>$ are the
eigenstates of the observables $A$ and $B$ that are measured at times
${t}_a$ and ${t}_b$ (referred to the internal clock) respectively, and
$|\psi({0})\>$ is the initial state of the system.

The correct two time correlations are obtained from the state
\eqref{psi} through the Born rule: the probability of obtaining the
$b$-th outcome at the measurement of $B$ when the $a$-th outcome was
obtained for $A$ is \begin{align} p(b|a)=|\<b|U_{{t}_b-{t}_a}|a\>|^2
\labell{cor}\;.
\end{align}
It follows from the joint probability
$p(a,b)=|\<b|U_{{t}_b-{t}_a}|a\>\<a|\psi({t}_a)\>|^2$ obtained from
the third integral, the probability $p(a)=|\<a|\psi({t}_a)\>|^2$ from
the second, and the Bayes rule $p(b|a)=p(a,b)/p(a)$.

\section{Experiment}
\begin{figure}[hbt] \begin{center}
\includegraphics[width=0.48\textwidth]{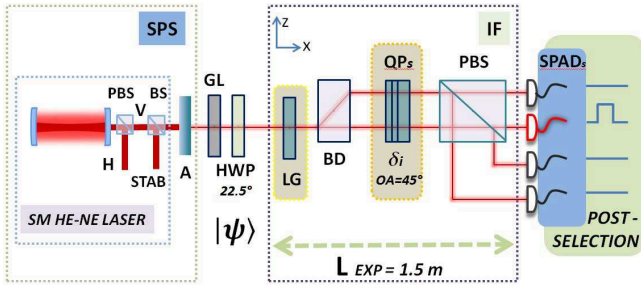}
\caption{Experimental setup: Two-time measurements in the PW mechanism. The SPS box shows the coherent source attenuated (A) at the photon-counting regime. A system composed by a polarizer (GL) and a half wave plate (HWP) oriented at $22.5^\circ$ pre-select the single photon state $|\Psi \>$. The IF box illustrates the two time
  measurements, operated by the two polarizing beam splitters BD (selecting the mode a=+1,-1)
  and PBS (selecting the mode b=+1,-1), respectively.  The blue boxes QPs represent different
  thicknesses of birefringent plates which perform the evolution of the photon by
  rotating its polarization: different thicknesses represent
  different time evolutions. At the output of the interferometer 4 SPADs connected to a coincidence electronic chain perform the single-photon detection.
  The birefringent plate LG in dashed box is used for the Leggett-Garg experiment only.}\label{f:setup}
\end{center}
\end{figure}

The experimental implementation of the ``timeless''
state \eqref{psi} takes advantage of a single narrow-band photon (Fig.~\ref{f:setup}), where its position
along the $x$ axis plays the role of the clock ($c$) degree of freedom, its
polarization acts as the system ($s$), and finally its position along the $z$ axis
implements the two memories ($m_1$, $m_2$).
A narrow-band attenuated coherent state (single cavity mode He-Ne Laser) is used as
source of single photons (SPS). In particular, the He-Ne laser is operated with a frequency stabilization feed-back control that balances the intensity of two lasing modes under the gain curve (with a nominal FWHM $\sim 1.5~GHz$). The tube length allows only two cavity modes at the output (with a nominal Free Spectral Range, FSR $\sim 1 ~GHz$).
Due to the fact that the polarization states of the two modes are orthogonal (i.e. one s-polarized and the other p-polarized), the single mode output is featured using a polarizing beamsplitter (SPS box in Fig.~\ref{f:setup}). By this way, one mode is redirected to a control photodetector, while the remaining mode passes through a second beamsplitter (Ratio 95\%-5\%) used for frequency stabilization.
The  linewidth  of  the individual  axial  mode emerging from the stabilized He-Ne, nominally with a narrow spectrum ($\sim kHz$), guarantees that the
clock (the photon's position) is, as requested by \eqref{psi}, in an
approximately uniform superposition of all times, at least over the
experimental time-scales: to fulfill this condition in the present experimental implementation, the bandwidth of the photon is requested to be smaller
than $300$ MHz to uniformly spread its position over an experimental
setup of size $\sim 1.5 m$.

  \begin{figure}[hbt] \begin{center}
\includegraphics[width=0.47\textwidth]{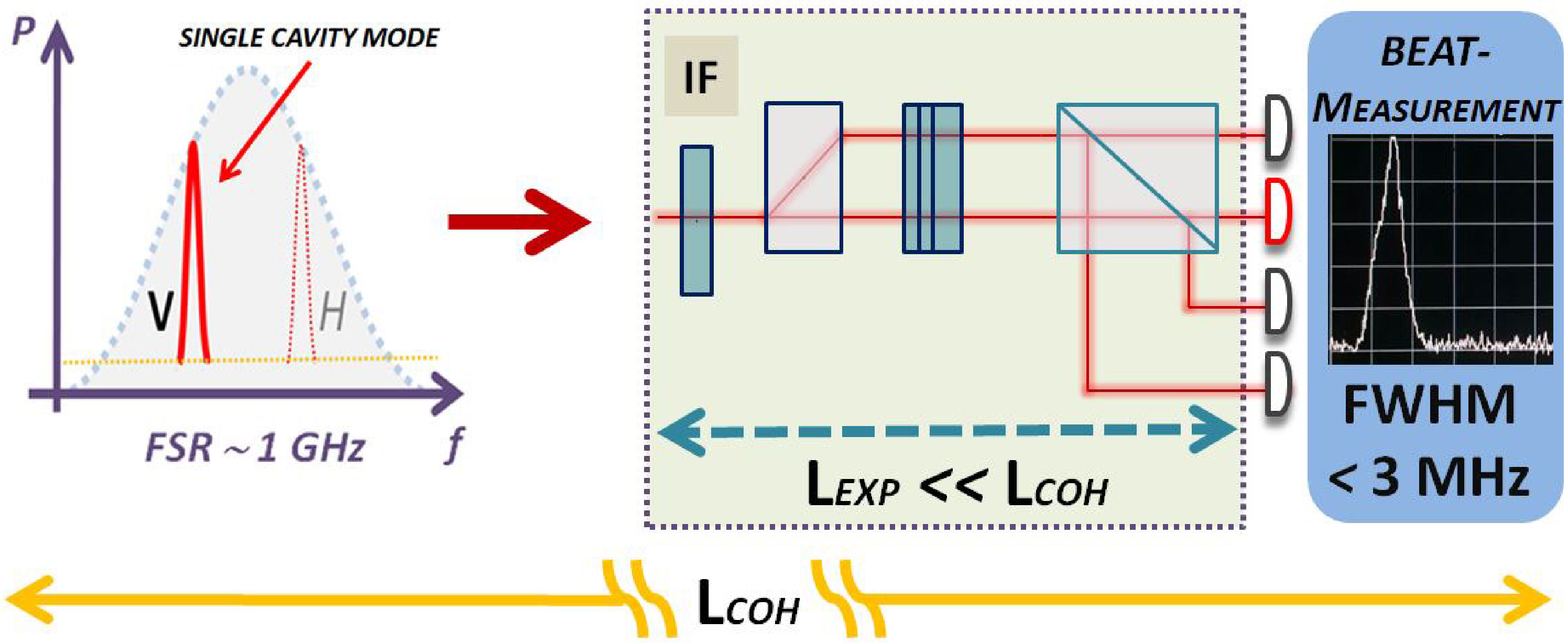}
\caption{Experimental setup: Super-Observer Mode. The SPS is an attenuate He-Ne laser operated with a frequency stabilization feed-back control that balances the intensity of two lasing modes under the gain curve ($\sim 1.5GHz$). The tube length allows only two cavity modes at the output (FSR $\sim 1 GHz$) with orthogonal polarization. After the suppression of one polarization mode (by a PBS), only one single cavity mode emerges from the source. The  linewidth  of this individual  axial  mode is evaluated by a beat measurement. The spectral analysis returned a bandwidth lower than 3 MHz, corresponding to a coherence length greater than 100 m.}\label{f:super-setup}
\end{center}
\end{figure}

After the attenuation (A) of the laser radiation at the single photon regime, the initial
polarization state of the photon (the system $s$) is selected with a
Glan-Thompson prism (GL) and a half-wave plate at $22.5^\circ$ (HWP) as a uniform
superposition of horizontal and vertical polarizations,
$|\psi(0)\>=\tfrac{1}{\sqrt{2}}\left(|H\>+|V\>\right)$. The Two-Time measurements are
performed by shifting the photon in the $z$ direction depending on its
polarization: the first (non-demolition) measurement is performed by the beam
displacer (BD) and the second by the polarization beam splitter
(PBS). This implements a von Neumann measurement
\cite{vonneumannbook}, where the system (polarization) is correlated
to a memory (the position along the $z$ axis). The BD and PBS
represent von Neumann's instantaneous unitary transformations
$U_{vn}$.  In accordance to Eq.~\eqref{psi}, these measurements are
performed at internal times $t_a$ and $t_b$: since the clock (internal
time) is represented by the photon's position along the $x$ direction,
the measurement times can be changed by moving BD and PBS along the
$x$ direction, which allows us to vary ${t}_a$ and ${t}_b$ at will.

\begin{figure}[tbp] \begin{center}
\includegraphics[width=0.42\textwidth]{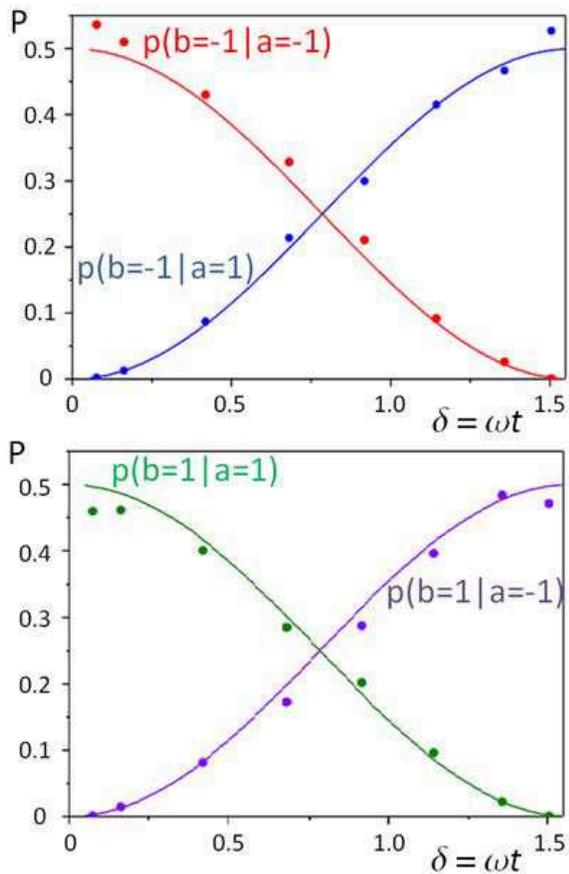}
\caption{Two-time
  correlation experiments. Graphs of conditional probabilities of
  polarization measurement outcomes as a function of the plate
  thickness (corresponding to the internal time evolution $t$). As
  expected, these probabilities are in good agreement with the
  conventional quantum description \eqref{correl} of the evolution
  from the internal observer's point of view.}
\label{f:results} \end{center} \end{figure}

To be more specific, the system's time evolution $U_t$ of Eq.~\eqref{psi} is implemented
through a set of birefringent quartz plates (QPs, with optical axes oriented at $45^\circ$) of various
optical thickness $\delta_i$, and positioned along the $x$ axis: varying
the thickness is equivalent to increase or decrease the optical
path length along the $x$ axis, namely to vary the interval
${t}_b-{t}_a$. Indeed, moving a plate to the left of BD corresponds to
decrease ${t}_a$, while moving it to the right of PBS corresponds to
increase ${t}_b$. Four Single-Photon Avalanche Detectors (SPADs) linked to a coincidence scheme with
$1$ ns time window are used as single particle detectors.  The ratios
of single counts revealed at each SPAD give the probability distributions
associated to the two-time measurement events. We also checked the
absence of coincidences between different pairs of detectors, which
corresponds to post-select the operation to the single photon
regime.  The experimental results for the conditional probabilities
(two-time correlations) are shown in Fig.~\ref{f:results}. Although in
this setup they have this meaning only from the internal observer's
point of view, they are in good agreement with the theoretical quantum
two-time correlations, as shown in Eq.~\eqref{cor}. In this case:

\begin{eqnarray}
p(b|a)=\left\{\begin{matrix}\cos^2[\omega({t}_b-{t}_a)]/2&
\mbox{ for }a=b\cr
\sin^2[\omega({t}_b-{t}_a)]/2&\mbox{ for }a\neq b\cr\end{matrix}\right.
\labell{correl}\;,
\end{eqnarray}
where $a,b=-1,1$ and $\omega=\delta_i/t_i$

From the external observer's point of view, we have to assess that the
global state is an eigenstate of the total Hamiltonian, namely that it
is stationary (with respect to an external clock). In other words, we
have to check that the clock is in a uniform coherent superposition,
being at different times coherently spread over the whole extent of the experiment (IF box in the pictures). This has been
implemented by measuring the bandwidth of the He-Ne laser emission (Fig.~\ref{f:super-setup}). To serve this purpose a beat signal has been measured superimposing the He-Ne Laser used in the experiment with a 632.816 nm Iodine-stabilized laser standard (INRiM 4/P frequency standard) \cite{TJQ}. The measured value is strictly below 3 MHz, returning a value for the coherence length of the photon over than 100 m \cite{MandelWolf}, confirming that the photon's position (clock) is
delocalized over the size of the present experiment (IF box length $\sim 1.5 m$). Moreover, an optical measurement of the coherence length of the photon has been performed by means of an unbalanced Michelson interferometer over a distance equal to $3m$ (distance imposed by spatial limitations in the laboratory), confirming the visibility of the fringes and that the coherence of the source is larger than the extension of the experiment.

\subsection*{Leggett-Garg}
Since we can access arbitrary two-time correlation measurements, we
can show that our experiment has a quantum behavior through a
violation of the Leggett-Garg inequality \cite{lg,indiani}.  This
inequality is constructed using an observable $Q(t)$ of a two-level
physical system, the photon's polarization in our case \cite{lgp}, with
$\left|H\right\rangle$, $\left|V\right\rangle$ corresponding to
$Q=+1,-1$ respectively.  The two-times correlation function is defined
as $C(t_i,t_j)=\<Q(t_i)Q(t_j)\>$ and can be obtained from the joint
probability $P_{ij}^{(t_i,t_j)}$ of obtaining the results $Q_i=Q(t_i)$
and $Q_j=Q(t_j)$ from the measurements at times $t_i$, $t_j$ :
\begin{equation}
 C({t_i,t_j})=\sum \limits_{Q_i, Q_j=\pm 1} Q_iQ_jP_{ij}^{(t_i,t_j)}.
\label{eq:C_fun}
\end{equation}
The Leggett-Garg inequality follows by supposing a  classical
realistic description of the system, which implies \cite{lg}
\begin{eqnarray}
&& -3\leq K_{3}\leq 1,
\label{eq:K13}
\\&&\mbox{ with }
K_{3}=C({t_1,t_2})+C({t_2,t_3})-C({t_1,t_3}),
\label{eq:K_fun}
\end{eqnarray}
for times $t_1<t_2<t_3$.  The inequality (\ref{eq:K13}) is a Bell-type
inequality and imposes realistic constraints on time-separated joint
probabilities. We choose $t_1$ as the initial time, i.e. $t_1=0$, and
assume that $t_{2}-t_{1}=t_{3}-t_{2}\equiv\Delta t$.

\begin{figure}[th] \begin{center}
\includegraphics[width=0.46\textwidth]{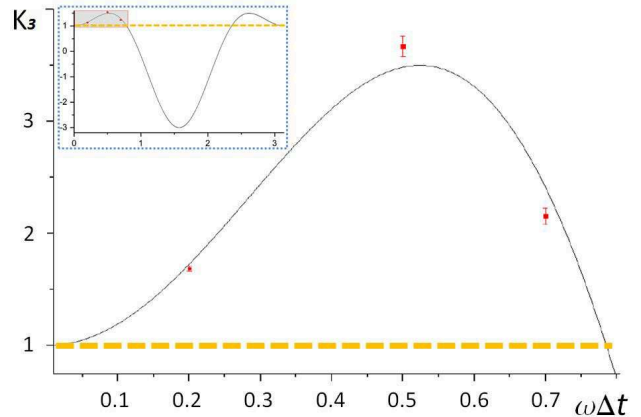}
\caption{Leggett-Garg function $K_3$ of Eq.~\eqref{eq:K_fun} for a
  two-level system as a function of measurement-time $\omega\Delta t$.
  The solid curve shows the theoretical $K_3$ of Eq.~\eqref{eq:K_fin},
  the red points are the  experimental results.}
\label{f:K_3} \end{center} \end{figure}

\begin{table}[h]
\centering
\label{my-label}
\begin{tabular}{|l|l|l|l|}
  \cline{1-4}
  $\Delta$t & Theory & Experiment&L-G violation in \\& & & standard deviation
    units  \\ \cline{1-4}
  $0.2$   & $1.159$  & $1.138 \pm 0.004 $ & $\approx 35 s.d.$ \\ \cline{1-4}
  $0.5$   & $1.499$  & $1.538 \pm 0.018$ & $\approx 30 s.d.$  \\ \cline{1-4}
  $0.7$   & $1.282$  & $1.238 \pm 0.018$ & $\approx 20 s.d.$   \\ \cline{1-4}
\end{tabular}
\caption{Comparison between the theoretical and experimental values of
  $K_3$ of Eq.~\eqref{eq:K_fun}. According to the Leggett-Garg
  inequality \eqref{eq:K13}, a value larger than one cannot come from
  a classical realistic system. }
\end{table}

To measure $C(t_i,t_j)$, we repeat the experiment varying the time
evolution by varying the thickness of the birefringent quartz plates
as described above. This implements the evolutions $U_t$ of the form
\begin{eqnarray}
  &&  U_{{t}=\delta/\omega} \left| H \right\rangle= \left| H \right\rangle\cos\delta +i\left|
    V \right\rangle\sin\delta\\&&
  U_{{t}=\delta/\omega} \left| V \right\rangle= i\left| H \right\rangle\sin\delta +\left|
    V \right\rangle\cos\delta ,
\label{eq:pol_evol}
\end{eqnarray}
where $\delta$ is the material's optical thickness.  Then the
conditional probability to find the photon in the state $\left| V
\right\rangle$ at time $t_1$ given that it was in $\left| V
\right\rangle$ at time $t_2$ is $P_{VV}^{(t_1,t_2)}=\tfrac{1}{2}
\cos^{2}\omega\Delta t$, in accordance to Eq.~\eqref{correl}.
Calculating also the probabilities $P_{HH}^{(t_1,t_2)}$,
$P_{HV}^{(t_1,t_2)}$, $P_{VH}^{(t_1,t_2)}$, we find
\begin{eqnarray}
  C({t_1,t_2})&=&
  \cos^2\omega\Delta t-\sin^2\omega\Delta t.
\label{eq:C_12}
\end{eqnarray}
The quantity $C({t_1,t_3})$ has the same form as $C({t_1,t_2})$, but
with a double optical thickness $\omega\Delta t \to 2 \omega\Delta t
$.  In order to measure $C(t_2,t_3)$, an additional birefringent
quartz plate (LG in Fig.~\ref{f:setup}) with an optical thickness $\omega\Delta t$ was introduced
before the $BD$ and a plate with the same optical thickness after
(refer to QPs in Fig.~\ref{f:setup}), obtaining
$C({t_2,t_3})=\cos^2\omega\Delta t-\sin^2\omega\Delta
t$. This configuration gives the ability to realize noninvasive
measurements at $t_2$ and projective measurement at $t_3$.
Substituting the values of $C(t_1,t_2)$, $C(t_2,t_3)$ and $C(t_1,t_3)$
into (\ref{eq:K_fun}) we find
\begin{equation}
 K_{3}=2\cos^2\omega\Delta t-2\sin^2\omega\Delta t-\cos^2 2\omega\Delta t+\sin^2 2\omega\Delta {t},
\label{eq:K_fin}
\end{equation}
whose maximum is for $\omega\Delta t =\pi/ 6$.  In
Fig.~\ref{f:K_3} and in Table I we compare this curve with the
measured values of $K_3$, obtained from the measured values of the
two-time correlations for three different values of $\Delta t$. The
measured values are in good agreement with the theoretical ones.

\section{Conclusions}

In summary in this paper we provided an experimental illustration of Page
and Wootters' quantum time mechanism that is able to describe multiple
two-time quantum correlation functions, giving us access to the
possibility of a test of the Leggett-Garg inequalities. In this scheme
the ``external'' observer sees a time-independent global state, in our
case a narrow-band single photon where the clock degree of freedom is
encoded in the photon's position, while the internal observer that
measures the position can track the flow of time.  In this framework we
tested a Leggett-Garg inequality, showing a violation from the
``internal'' observer point of view.  Our results, showing the Page
Wootters scheme at work, pave the way for further studies, addressed
to understand time in quantum mechanics.

\subsection*{Acknowledgments}

LM acknowledges the FQXi foundation for financial support in the
``Physics of what happens'' program. The authors acknowledge the ``COST Action MP1405 QSPACE'' for financial support.
The authors would like to thank Dr. Marco Bisi, for his support in the
beat signal measurement operated with INRiM 4/P frequency standard,
Dr. Marco Pisani and Dr. Massimo Zucco for useful discussions, F. Veneziano for collaboration.


\begin{references}
\bibitem{qgt} A. Ashtekar, Studies in History and Philosophy of Science Part B: Studies in History and Philosophy of Modern Physics {\bf 52}, Part A, 69–74 (2015);  E. Anderson, Ann. Phys. 524, 757 (2012); C. Rovelli, Phys. Rev. D {\bf 43}, 442 (1991); W. Unruh and R. Wald, Phys. Rev. D {\bf 40} 2598 (1989), R. Gambini et al., Phys. Rev. D {\bf 79}, 041501 (2009); D.T. Pegg,  J. Phys. A {\bf 24}, 3031 (1991).
\bibitem{paw}D.N. Page and W.K. Wootters, Phys. Rev. D {\bf 27}, 2885 (1983).
\bibitem{pauli} J. Leon, L. Maccone, 
  arXiv:1705.09212 (2017).
\bibitem{morse}P. McCord Morse, H. Feshbach, {\em Methods of
    Theoretical Physics, Part I} (McGraw-Hill, 1953), Chap. 2.6;  H. D.
  Zeh,
  Phys. Lett. A {\bf 116}, 9 (1986).
\bibitem{altri} H.D.  Zeh, {\em Time in quantum
    theory},
  http://www.rzuser.uni-heidelberg.de/$\sim$as3/TimeInQT.pdf; T.
  Banks, 
  Nucl. Phys. B {\bf 249}, 332 (1985); R. Brout, 
  Found. Phys. {\bf 17}, 603 (1987); R. Brout, G. Horwitz, D. Weil,
  Phys. Lett. B
  {\bf 192}, 318 (1987); R. Brout, 
  Z. Phys. B {\bf 68}, 339 (1987); V. Vedral,
  arXiv:1408.6965 .
\bibitem{ak}Y. Aharonov, T. Kaufherr, 
Phys. Rev. D {\bf 30}, 368 (1984).
\bibitem{kuchar}K.V. Kucha\u r, ``Time and interpretations of quantum
  gravity'', Proc. 4th Canadian Conference on General Relativity and
  Relativistic Astrophysics, eds. G. Kunstatter, D.  Vincent, and J.
  Williams (World Scientific, Singapore, 1992), pg.~69-76.
\bibitem{unruhwald}W.G. Unruh, R.M. Wald, 
Phys. Rev. D {\bf 40}, 2598 (1989).
\bibitem{sorkin} R.D. Sorkin, 
Int. J. Theor.  Phys. {\bf 33}, 523
  (1994).
\bibitem{qtime}V. Giovannetti, S. Lloyd, L. Maccone, 
  Phys. Rev. D, {\bf 92}, 045033 (2015).
\bibitem{esperimentotorino} E. Moreva, G. Brida, M. Gramegna, V.
  Giovannetti, L. Maccone, M. Genovese, 
  Phys. Rev. A {\bf 89}, 052122 (2014).
\bibitem{TJQ} T.J.   Quinn,   
Metrologia {\bf 40}, 103-133 (2003).
\bibitem {MandelWolf} L. Mandel, E.  Wolf, Optical Coherence and Quantum Optics (Cambridge University Press, 1995).
\bibitem{lg} A.J. Leggett and A. Garg, 
Phys. Rev. Lett. {\bf 54}, 857   (1985).
\bibitem{indiani}D. Gangopadhyay, A. Sinha Roy, arXiv:1608.06070 .
\bibitem{pagereply}D.N. Page, ``Clock time and entropy'', in {\em
    Physical Origins of Time Asymmetry}, eds. J.J. Halliwell, et al.,
  (Cambridge Univ. Press, 1993), arXiv:gr-qc/9303020 .
\bibitem{vonneumannbook} J. von Neumann, {\em Mathematical
    Foundations of Quantum Mechanics} (Princeton Univ. Press, 1955).
    \bibitem{lgp} A. Avella, et al., arXiv:1706.02120; M. Goggin et al., 
    Proc. Natl. Acad. Sci. U. S. A. {\bf 108}, 1256 (2011); Jin-Shi Xu et al., 
    Scientific Report {\bf 1}, 101 (2011).





\end{references}
\end{document}